\documentclass[11pt,a4]{article}
\begin{document}
\begin{titlepage}
\begin{flushright}
\parbox{4cm}{
\baselineskip=12pt
 TMUP-HEL-9904\\
 September, 1999\\
\hspace*{1cm}}
\end{flushright}
\vfil
\begin{center}
 \Large
 Physical auxiliary field in supersymmetric QCD
 with explicit supersymmetry breaking
\end{center}
\vfil
\begin{center}
 \large
 Noriaki Kitazawa
\end{center}
\begin{center}
 \small\it
 Department of Physics, Tokyo Metropolitan University,\\
 Hachioji, Tokyo 192-0397, Japan
\end{center}
\vfil
\begin{abstract} It is shown that
 the auxiliary field in the low-energy effective theory of the
 supersymmetric QCD (SU$(N_c)$ gauge symmetry with flavors $N_f < N_c$)
 can be understood as a physical degree of freedom,
 once the supersymmetry is explicitly broken.
Although the vacuum expectation value of the auxiliary field
 is just a measure of the supersymmetry breaking
 in the perturbative treatment of the supersymmetry breaking,
 it can be the vacuum expectation value of the quark bilinear operator
 in the non-perturbative treatment of the supersymmetry breaking.
We show that the vacuum expectation value remains finite
 in the limit of the infinite supersymmetry-breaking mass of the squark.
We have to take the large $N_c$ limit simultaneously
 to keep the low-energy effective K\"ahler potential
 being in good approximation.
\end{abstract}
\end{titlepage}
\newpage

\section{Introduction}

It is natural to attempt extracting some non-perturbative information
 on the non-supersymmetric gauge theory from the supersymmetric one
 by introducing the explicit supersymmetry breaking,
 since some non-perturbative effects can be reliably evaluated
 in the latter theory\cite{MV,ASPY,DMS,EHS,MW,ADKM,AM,AMZ}.
In case of studying QCD using the supersymmetric QCD (SQCD)
 \footnote{
  In this letter we use the word QCD
   to indicate general vector-like SU$(N_c)$ gauge theories
   with $N_f$ flavors}
 there is a problem of physical degrees of freedom\cite{SS1}.
In the low-energy effective theory of SQCD
 there is no physical field which can correspond to
 the low-energy effective field in QCD.
All the physical low-energy effective fields in SQCD
 describe the bound states which include at least one squark,
 and there is no effective field corresponding to the bound state
 whose constituents are quarks only.
The auxiliary field in the effective ``meson'' superfield
 seems to correspond to the bound sates of two quarks (mesons)\cite{SS1},
 but it is not clear if the auxiliary field really
 can be a physical degree of freedom
 \footnote{
  The low-energy effective scalar field with mass dimension three,
   which is the same mass dimension of the auxiliary field,
   is successfully utilized in Ref.\cite{SS2}.}\cite{KS}.

In this note we show that
 the auxiliary component of the effective superfield
 in SQCD with $N_f < N_c$
 can be understood as a physical degree of freedom.
We introduce
 the supersymmetry-breaking squark mass in the ultraviolet theory,
 and construct its low-energy effective theory
 following the method in Refs.\cite{S,ADS,PR}.
It gives the scalar potential which is described by two scalar fields,
 the first and auxiliary components of the effective superfield,
 and the vacuum expectation values (VEV) of these fields
 can be obtained.
If we naively take the limit of the infinite squark mass,
 the VEV of the auxiliary field diverges.
This result does not mean the unphysical nature of the field,
 but the bad approximation for the effective K\"ahler potential.
If we take large $N_c$ limit simultaneously,
 the VEV of the auxiliary field remains finite,
 and the auxiliary field can be understood
 as a physical degree of freedom corresponding to the meson in QCD.
Large $N_c$ limit is required
 to keep the effective K\"ahler potential in good approximation.

\section{SQCD with explicit supersymmetry breaking}

The Lagrangian of the supersymmetric SU$(N_c)$ gauge theory
 ($N_c > 2$) with flavors $N_f < N_c$ is
\begin{equation}
 {\cal L} =
 - \int d^4 \theta K
 + \left\{
    {1 \over 2} {\rm tr} \int d^2 \theta W^{\dot{\alpha}} W_{\dot{\alpha}}
  + {\rm h.c.}
   \right\}
\end{equation}
 and
\begin{equation}
 K = Q^{\dag}_i e^{-2gV} Q^i + {\bar Q}^{\dag i} e^{2gV^T} {\bar Q}_i,
\label{K-susy}
\end{equation}
 where $Q^i$ and ${\bar Q}_i$
 are quark chiral superfields with flavor index $i$,
 $V$ and $W^{\dot{\alpha}}$ are the gluon vector superfield
 and its field strength chiral superfield, respectively,
 and $g$ is the gauge coupling (see Ref.\cite{K} for notation).
The supersymmetry-breaking mass for squarks can be introduced as
\begin{equation}
 {\cal L}_{m} = - \int d^4 \theta X K,
\label{K-susy-br}
\end{equation}
 where $X$ is an interpolating vector superfield
 whose highest component has a VEV
 $\langle X \rangle = \theta^2 {\bar \theta}^2 m^2$.
In the infinite mass limit
 the theory becomes non-supersymmetric QCD with an adjoint fermion.

It is known that
 the low-energy effective field of this theory with $m=0$
 is the gauge singlet chiral superfield
\begin{equation}
 M^i{}_j \sim Q^i {\bar Q}_j.
\label{eff-field}
\end{equation}
The correspondence of its component fields
 to the composite operators in the fundamental theory is
\begin{eqnarray}
 A_M &\sim& A_Q A_{\bar Q}
  \qquad\qquad\quad \mbox{scalar component}, \\
 \psi_M &\sim& \psi_Q A_{\bar Q} + A_Q \psi_{\bar Q}
  \quad \mbox{fermion component}, \\
 F_M &\sim& - \psi_Q \psi_{\bar Q}
  \qquad\quad\quad \ \mbox{auxiliary component},
\label{corresp}
\end{eqnarray}
 where the auxiliary fields of quarks are integrated out
 by using the equation of motion.
Note that
 the auxiliary field $F_M$ corresponds to the quark bilinear operator.
The low-energy effective theory is described as
\begin{equation}
 {\cal L}^{\rm eff}
 = - \int d^4 \theta K^{\rm eff}
   + \left\{
      \int d^2 \theta W_{\rm dyn} + {\rm h.c.}
     \right\},
\end{equation}
 where $W_{\rm dyn}$ is the dynamically generated superpotential
\begin{equation}
 W_{\rm dyn}
 = (N_c-N_f) \left(
              {{\Lambda^{3 N_c - N_f}} \over {\det M}}
             \right)^{1 \over {N_c-N_f}}
\end{equation}
 and $\Lambda$ is the scale of dynamics\cite{S}.
It is very difficult
 to obtain the exact effective K\"ahler potential.
We use
\begin{equation}
 K^{\rm eff} = 2 {\rm tr} \sqrt{M^{\dag} M}
\label{K-eff-susy}
\end{equation}
 which is effective in the weak coupling limit
 and describes the theory on the classical flat direction
 \cite{ADS,PR}.

The supersymmetry breaking effect due to $X$
 can be introduced to the effective theory by including the terms
 which are the general functions of $X$ and $M$.
Since $X$ is a vector superfield and we can consider as $X^2=0$,
 only the K\"ahler potential has correction of the form
\begin{equation}
 K_m^{\rm eff} = X F(M^{\dag}, M)
\label{K-eff-susy-br}
\end{equation}
 in the low-energy limit, where $F$ is a real function.
We estimate the function $F$ as follows.

Since the K\"ahler potential of Eq.(\ref{K-eff-susy})
 is obtained by a simple replacement of Eq.(\ref{K-susy})
 based on the identification of Eq.(\ref{eff-field}),
 we may be able to do the same replacement
 in Eq.(\ref{K-susy-br}).
But the replacement itself
 must be modified by the supersymmetry breaking effect.
From the condition of the flat direction, we have\cite{ADS}
\begin{equation}
 ( {\bar Q}^{\dag} {\bar Q} )^i{}_k
 ( {\bar Q}^{\dag} {\bar Q} )^k{}_j
 = ( {\bar Q}^{\dag} Q^{\dag} )^i{}_k
   ( Q {\bar Q} )^k{}_j.
\end{equation}
In the supersymmetric case
 the right hand side is simply replaced by $M^{\dag}M$,
 but now we should have
\begin{equation}
 ( {\bar Q}^{\dag} {\bar Q} )^i{}_k
 ( {\bar Q}^{\dag} {\bar Q} )^k{}_j
 = M^{\dag i}{}_k M^k{}_j + X G(M^{\dag},M)^i{}_j,
\label{op-match}
\end{equation}
 where $G$ is a real function of $M^{\dag}$ and $M$.
This relation results
\begin{equation}
 {\rm tr} \left( {\bar Q}^{\dag} {\bar Q} \right)
 = {\rm tr} \sqrt{M^{\dag} M}
 + {1 \over 2} X {\rm tr}
   \left( (M^{\dag} M)^{-{1 \over 2}} G(M^{\dag},M) \right).
\end{equation}
Therefore, we have
\begin{equation}
 F(M^{\dag},M)
 = 2 {\rm tr} \sqrt{M^{\dag} M}
 + {\rm tr} \left( (M^{\dag} M)^{-{1 \over 2}} G(M^{\dag},M) \right)
\label{susy-br-cont}
\end{equation}
 in the weak coupling limit.
The result of this note indicates that
 the first term dominates the second term in large $N_c$ limit.
Here, we simply neglect the second term.
The effective theory which we are going to analyze is
\begin{equation}
 {\cal L} =
 - \int d^4 \theta \left( 1 + X \right) K^{\rm eff}
 + \left\{
    \int d^2 \theta W_{\rm dyn} + {\rm h.c.}
   \right\}.
\end{equation}

We can obtain the effective scalar potential from this Lagrangian.
For simplicity, we assume that
 the VEV of the effective field
 is flavor diagonal: $M^i{}_j = \Phi \cdot {\bf 1}^i{}_j$.
The scalar potential
 is described by two scalar fields, $A_\Phi$ and $F_\Phi$,
 the scalar and auxiliary components of the superfield $\Phi$,
 respectively.
\begin{equation}
 V =
 - {{N_f} \over 2}
    {{F_\Phi^{\dag} F_\Phi} \over \sqrt{A_\Phi^{\dag} A_\Phi}}
 + 2 N_f m^2 \sqrt{A_\Phi^{\dag} A_\Phi}
 + \left\{
    N_f {{F_\Phi} \over {A_\Phi}}
         \left(
          {{\Lambda^{3N_c-N_f}} \over {A_\Phi^{N_f}}}
         \right)^{1 \over {N_c-N_f}}
    + {\rm h.c.}
   \right\}.
\end{equation}
If we perturbatively solve the stationary conditions,
 $\partial V / \partial A_\Phi = 0$
 and $\partial V / \partial F_\Phi = 0$,
 the VEV of the auxiliary field is simply proportional to $m^2$,
 and it is merely a measure of the supersymmetry breaking
 (some regulator interaction, $W_{\rm reg} = \det M / \mu^{2N_f-3}$,
 for example, must be introduced to have a finite VEV of $A_\Phi$
 in the supersymmetric limit).
We can easily solve the stationary conditions without any approximation.
The result is
\begin{eqnarray}
 A_\Phi
  &=& \left(
       {{\Lambda^{3N_c-N_f}} \over {(Cm)^{N_c-N_f}}}
      \right)^{1 \over {N_c}},
\\
 F_\Phi
  &=& 2 C^{{N_f} \over {N_c}}
       \left(
        m^{N_f} \Lambda^{3N_c-N_f}
       \right)^{1 \over {N_c}},
\end{eqnarray}
 where $C \equiv \sqrt{(N_c-N_f)/(N_c+N_f)}$.

\section{Decoupling limit and VEV of the auxiliary field}

Before taking the limit of $m \rightarrow \infty$,
 the scale of dynamics at low energies, $\Lambda'$,
 must be specified.
Since we take $m$ very large
 in comparison with any other scales of dynamics,
 the matching condition of the running gauge coupling at one-loop
 can be utilized.
The scale $\Lambda$
 can be described by the scale $\Lambda'$ and $m$ as
\begin{equation}
 \Lambda
 = \Lambda' \left( {{\Lambda'} \over {m}} \right)^{{b'-b} \over b},
\label{scale-match}
\end{equation}
 where $b \equiv 3N_c-N_f$ and $b' \equiv 3N_c-{2 \over 3}N_f$.
Then, we have
\begin{eqnarray}
 A_\Phi
  &=& C^{-{{N_c-N_f} \over {N_c}}} \left( \Lambda' \right)^2
      \left( {m \over {\Lambda'}} \right)^{-1+{{2N_f} \over {3N_c}}},
\label{scalar}\\
 F_\Phi
  &=& 2 C^{{N_f} \over {N_c}} \left( \Lambda' \right)^3
      \left( {m \over {\Lambda'}} \right)^{{2N_f} \over {3N_c}}.
\label{auxiliary}
\end{eqnarray}
In $m \rightarrow \infty$ limit with fixed $\Lambda'$,
 $F_\Phi$ diverges, though $A_\Phi$ vanishes as expected.
This is
 because of a truncation in the effective K\"ahler potential
 in the previous section.
We show that
 we can obtain a convincing result in large $N_c$ limit,
 which indicates that the first term in Eq.(\ref{susy-br-cont})
 dominates the second term in large $N_c$ limit.

To take a correct large $N_c$ limit
 the scales of dynamics, $\Lambda$ and $\Lambda'$,
 should not depend on $N_c$.
This requires that
\begin{equation}
 {{\Lambda'} \over \Lambda}
 = \left( {m \over {\Lambda'}} \right)^{{b'-b} \over b}
 \rightarrow \left( {m \over {\Lambda'}} \right)^{{N_f} \over {9N_c}}
\end{equation}
 should be a finite value.
If we take the finite value as $\beta^{1/6}$
 ($\beta > 1$ because of $\Lambda' > \Lambda$),
 we have
\begin{eqnarray}
 A_\Phi
  &\rightarrow&
  \left( \Lambda' \right)^2 \beta \ \beta^{-{{3N_c} \over {2N_f}}}
  \rightarrow 0,
\\
 F_\Phi &\rightarrow& 2 \left( \Lambda' \right)^3 \beta
\end{eqnarray}
 in large $m$ and $N_c$ limit.
The uncertainty of the value of $\beta$
 corresponds to the ambiguity in the present determination
 of the scale of dynamics at low energies.
Now we have a finite value of $F_\Phi$ and vanishing $A_\Phi$
 which is required to have non-decoupling quarks.
This result shows that
 the auxiliary field can be understood
 as a physical degree of freedom.

We note that
 the K\"ahler potential of Eq.(\ref{K-eff-susy})
 is in good approximation,
 since the value of the gauge coupling
 in the supersymmetric theory at the scale $\Lambda'$
 which is the typical scale of the low-energy physics
 is small.
\begin{equation}
 \alpha(\Lambda')
 = {{2\pi} \over {(3N_c-N_f) \ln(\Lambda'/\Lambda)}}
 \rightarrow {{2\pi} \over {3N_c \ln \beta^{1/6}}}
\end{equation}
We also note that $F_\Phi$ diverges in this large $m$ and $N_c$ limit,
 if we assume the naive effective K\"ahler potential
 $K^{\rm eff}_{\rm naive} \propto {\rm tr} (M^{\dag} M) / \Lambda^2$.

The next question is
 which degree of freedom in the underlying theory
 corresponds to the auxiliary field.
As described in Eq.(\ref{corresp}),
 it is natural to consider that
 the quark bilinear operator corresponds to the auxiliary field,
 though the large $N_c$ scaling is different
 ($\langle \psi_Q \psi_{\bar Q} \rangle \sim N_c$).
This means that
 the coefficient which should be determined by the dynamics
 depends on $N_c$.
Namely,
\begin{equation}
 F_M{}^i{}_j \propto - {1 \over {N_c}} \psi_Q^i \psi_{{\bar Q} j}
             \propto {1 \over \sqrt{N_c}} \Sigma^i{}_j,
\end{equation}
 where $\Sigma$ is the complex scalar field
 in the linear $\sigma$-model as the low-energy effective theory
 of QCD with an adjoint fermion.

\section{Conclusion}

It has been shown that
 the auxiliary component of the low-energy effective chiral superfield
 in the supersymmetric QCD ($N_f < N_c$)
 with explicit supersymmetry breaking
 can be understood as a physical degree of freedom.
Supersymmetry was explicitly broken by the squark mass $m$,
 and the large $m$ limit was considered
 to obtain the non-supersymmetric theory: QCD with an adjoint fermion.
Large $N_c$ limit was required
 to keep the effective K\"ahler potential in good approximation.
The finite vacuum expectation value
 of the auxiliary field was obtained in the limit,
 which means that the auxiliary field can be a physical degree of freedom
 in the non-supersymmetric theory.
It can be identified
 to the quark bilinear operator with a coefficient of $1/N_c$ scaling
 or the meson field with a coefficient of $1/\sqrt{N_c}$ scaling.

\vspace{1cm}
The author would like to thank Francesco Sannino for helpful discussions.
The author also would like to thank
 the physics department of Yale university for the hospitality, 
 where most of this work was done.
This work was supported in part
 by the Grant-in-Aid for Scientific Research
 from the Ministry of Education, Science and Culture of Japan
 under contract \#11740156.


\begin{thebibliography}{99}
\bibitem{MV}
 A.Masiero and G.Veneziano, Nucl. Phys. B249 (1985) 593.
\bibitem{ASPY}
 O.Aharony, J.Sonnenshein, M.E.Peskin and S.Yankielowicz,
 Phys. Rev. D52 (1995) 6157.
\bibitem{DMS}
 E.D'Hoker, Y.Mimura and N.Sakai, Phys. Rev. D54 (1996) 7724.
\bibitem{EHS}
 N.Evans, S.D.H.Hsu and M.Schwetz,
 Phys. Lett. 404B (1997) 77; Nucl. Phys. B484 (1997) 124.
\bibitem{MW}
 S.P.Martin and J.D.Wells, Phys. Rev. D58 (1998) 115013.
\bibitem{ADKM}
 L.Alvarez-Gaume, J.Distler, C.Kounnas and M.Marino,
 Int. J. Mod. Phys. A11 (1996) 4745.
\bibitem{AM}
 L.Alvarez-Gaume and M.Marino, Int. J. Mod. Phys. A12 (1997) 975.
\bibitem{AMZ}
 L.Alvarez-Gaume, M.Marino and F.Zamora,
 Int. J. Mod. Phys. A13 (1998) 403; {\it ibid} A13 (1998) 1847.
\bibitem{SS1}
 F.Sannino and J.Schechter, Phys. Rev. D57 (1998) 170.
\bibitem{SS2}
 F.Sannino and J.Schechter, Phys. Rev. D60 (1999) 056004.
\bibitem{KS}
 N.Kitazawa and F.Sannino, hep-th/9802017; hep-th/9803225.
\bibitem{S}
 N.Seiberg, Phys. Rev. D49 (1994) 6857.
\bibitem{ADS}
 I.Affleck, M.Dine and N.Seiberg, Nucl. Phys. B256 (1985) 557.
\bibitem{PR}
 E.Poppitz and L.Randall, Phys. Lett. 336B (1994) 402.
\bibitem{K}
 N.Kitazawa, hep-ph/9901257.
\end{thebibliography}
\end{document}